\documentclass[runningheads]{svmult}

\usepackage{makeidx}   
\usepackage{graphicx}  
\usepackage{subeqnar}  
\usepackage{multicol}  
\usepackage{physprbb}  
\makeindex             

\begin{document}
\title*{Starbursts: Triggers and Evolution}
\toctitle{Starbursts: Triggers and Evolution}
%
%
\titlerunning{Starbursts: Triggers and Evolution}
%
%
\author{Shardha Jogee\inst{1}
}
%
\authorrunning{Shardha Jogee}
%
%
\institute{
Division of Physics, Mathematics, and Astronomy, MS 105-24, 
California Institute of Technology, Pasadena, CA 91125
}
\maketitle              
\begin{abstract}
Why do the circumnuclear (inner 1--2\,kpc) regions of spirals
show vastly different star formation rates (SFR) even if they
have a comparable molecular gas content? 
Why do some develop starbursts  which are intense short-lived 
(t\,$\ll$\,1\,Gyr) episodes of star formation  characterized
by a high star formation rate per  unit mass of molecular gas
(SFR/M$_{\mathrm{H2}}$), which I   
refer to as  star formation efficiency (SFE)\@.  
I address these questions  using  
high resolution (2$''$ or 100--200\,pc) CO  (J=1$\rightarrow$0) 
observations from the Owens Valley Radio Observatory,
optical  and NIR  images, 
along with published radio continuum (RC)  and Br$\gamma$ data.
The sample of eleven galaxies  includes  the brightest nearby 
starbursts comparable to M82 and control non-starbursts.
More detailed results are in \cite{J99} and \cite{JKS01}.

\end{abstract}

\section {External Disturbances and Large-Scale Stellar Bars}

The sample galaxies   have developed large molecular gas  reservoirs 
of several$\,\times\,$10$^{8}$ to several$\,\times\,$10$^{9}$\,M$_\odot$ in 
the inner kpc radius,  assuming  a standard  CO-to-H$_2$ 
conversion factor.  As shown  in Fig.~1, 
\it \rm 
the circumnuclear SFR per unit mass of  molecular gas 
spans more than an order of magnitude 
for a given  circumnuclear molecular gas content. 
\rm

\begin{figure}[h]
\begin{center}
\includegraphics[width=1.9in]{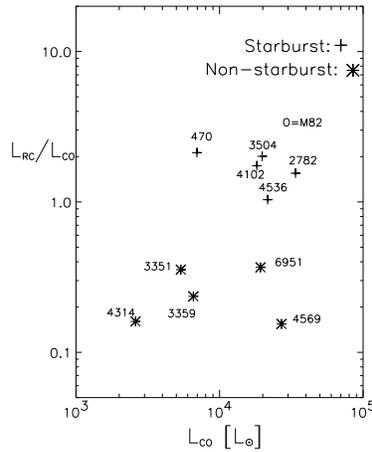}
\end{center}
\vspace{-0.2in}
\caption[]
{ The sample galaxies are shown.
L$_{\mathrm{RC}}$ is the RC luminosity at 1.5 GHz \cite{CHSS90}
and 
L$_{\mathrm{CO}}$ is the single dish CO luminosity \cite{Yetal95},
both measured in the   central $45''$\@.
}
\end{figure}   


\begin{figure}[ht]
\vspace{0.4in}
\begin{center}
\includegraphics[width=.8\textwidth]{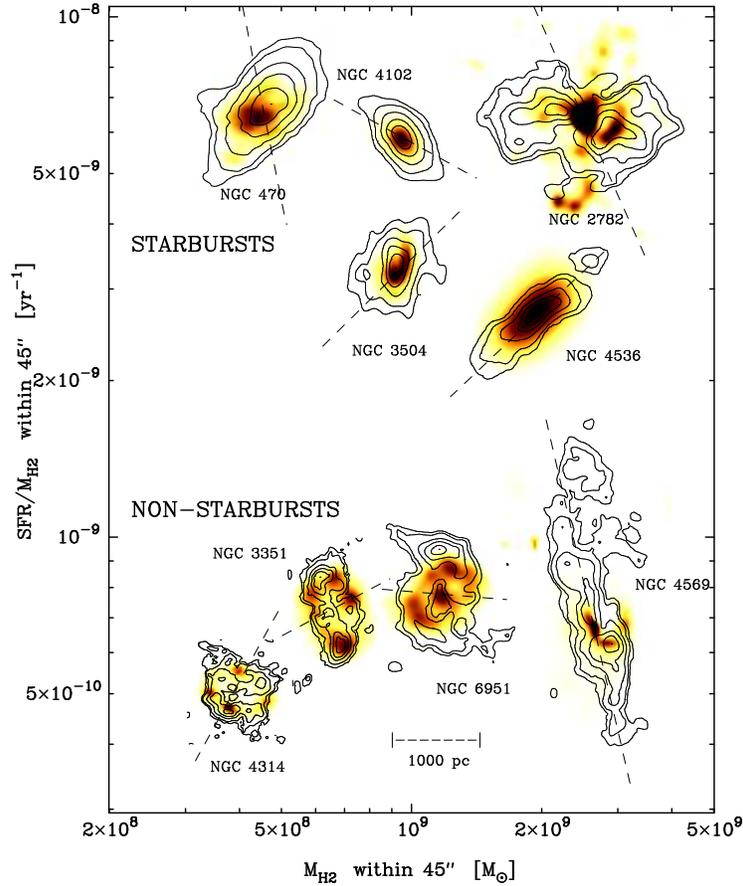}
\end{center}
\vspace{-0.1in}
\caption[]{
In the SFR/M$_{\mathrm{H2}}$ vs.~M$_{\mathrm{H2}}$ plane, 
the CO  intensity  (contours) is  overlaid on the  
star formation  (greyscale), as traced by RC in NGC\,4102, NGC\,2782, 
and NGC\,6951, 
and by H$\alpha$  in the others. 
The dotted line is  the P.A. of the  large-scale stellar bar/oval. 
The synthesized CO beam is typically 100--200\,pc.
}
\end{figure}

A spontaneously or tidally induced  m\,=\,2  instability such 
as a   large-scale  stellar bar  and 
minor  mergers/interactions can help
to drive gas  towards the inner kpc \cite{HM95}.  
In   optical and NIR images, all our sample galaxies 
show a large-scale stellar bar/oval whose position angle 
is marked on Fig.~2.
The  stellar bar may have been recently tidally triggered 
in NGC\,3359 which has  a steep abundance gradient 
along the bar \cite{MR95}, and in NGC\,4569 which 
has a warped disk, an asymmetric bar and disturbed CO 
properties \cite{J99}.
While the  galaxies in our sample are not major mergers, 
all of them except for NGC\,6951, show 
evidence for recent  tidal interactions or  mergers with 
mass ratios ranging from minor (1:10) to intermediate (1:4). 
NGC\,2782 and NGC\,470, which have  the  
largest mass-ratio interactions, are starbursts. 

The presence of a large-scale bar in starbursts and non-starbursts 
alike suggests that 
\it \rm  
the starburst lifetime is short with respect to the 
timescale for bar destruction. 
\rm
Thus,  within a given barred potential, 
star formation can change from  an inefficient pre-burst 
phase in the early stages of bar-driven inflow, to 
a circumnuclear starburst, 
to a post-starburst after star 
formation rapidly consumes gas near the center 
in a  few\,$\times\,$10$^{8}$\,years.
The bar itself may be weakened or dissolve over timescales $>$\,1\,Gyr 
due to the development of a high  central  mass concentration 
(e.g., \cite{HN90}; \cite{FB93})

\section{Morphology of the Circumnuclear Molecular Gas}

What is the circumnuclear CO morphology and how does it 
relate to the properties of the barred potential? 
The molecular gas shows a wide variety 
of  morphologies (Fig.~2) 
ranging from   relatively axisymmetric annuli or disks 
(starbursts NGC\,4102, NGC\,3504, NGC\,4536,   and 
non-starbursts NGC\,4314), elongated double-peaked  and spiral  
morphologies (starburst NGC\,2782 and non-starbursts NGC\,3351 and  
NGC\,6951)  to extended distributions elongated along the 
large-scale bar (non-starburst NGC\,4569). 
In  NGC\,4569, the gas extends out to a large 
(2\,kpc) radius, at a similar P.A. as the large-scale stellar bar 
(Fig.~2), and shows complex   non-circular motions. 
The   optical, NIR, and CO properties of 
NGC\,4569 suggest it  is in the early stages of bar-driven/tidally-driven 
inflow of gas towards the inner kpc. 
In the other  galaxies, the gas distribution is less 
extended,  and in many systems it is concentrated inside the outer 
inner Lindblad resonance (ILR)  of the large-scale bar.
As shown in Fig.~3, both starbursts and non-starbursts host 
ILRs. In the sample, the bar pattern speed 
$\Omega_{\mathrm{p}}> 40$--115\,km~s$^{-1}$~kpc$^{-1}$, 
the  radius of the outer ILR is typically $>$\,500\,pc, and the 
radius of the inner IILR  $<$\,300\,pc. 
Note that  in NGC\,2782 and NGC\,470 which 
are claimed to host nuclear stellar bars \cite{JKS99,FWRB96},
there is  a strong misalignment ($\ge$\,40\,$^\circ$) between the 
CO distribution and both the major axis and  minor axis of the  
large-scale stellar bar/oval. 

\begin{figure}[h]
\begin{center}
\includegraphics[width=2.0in]{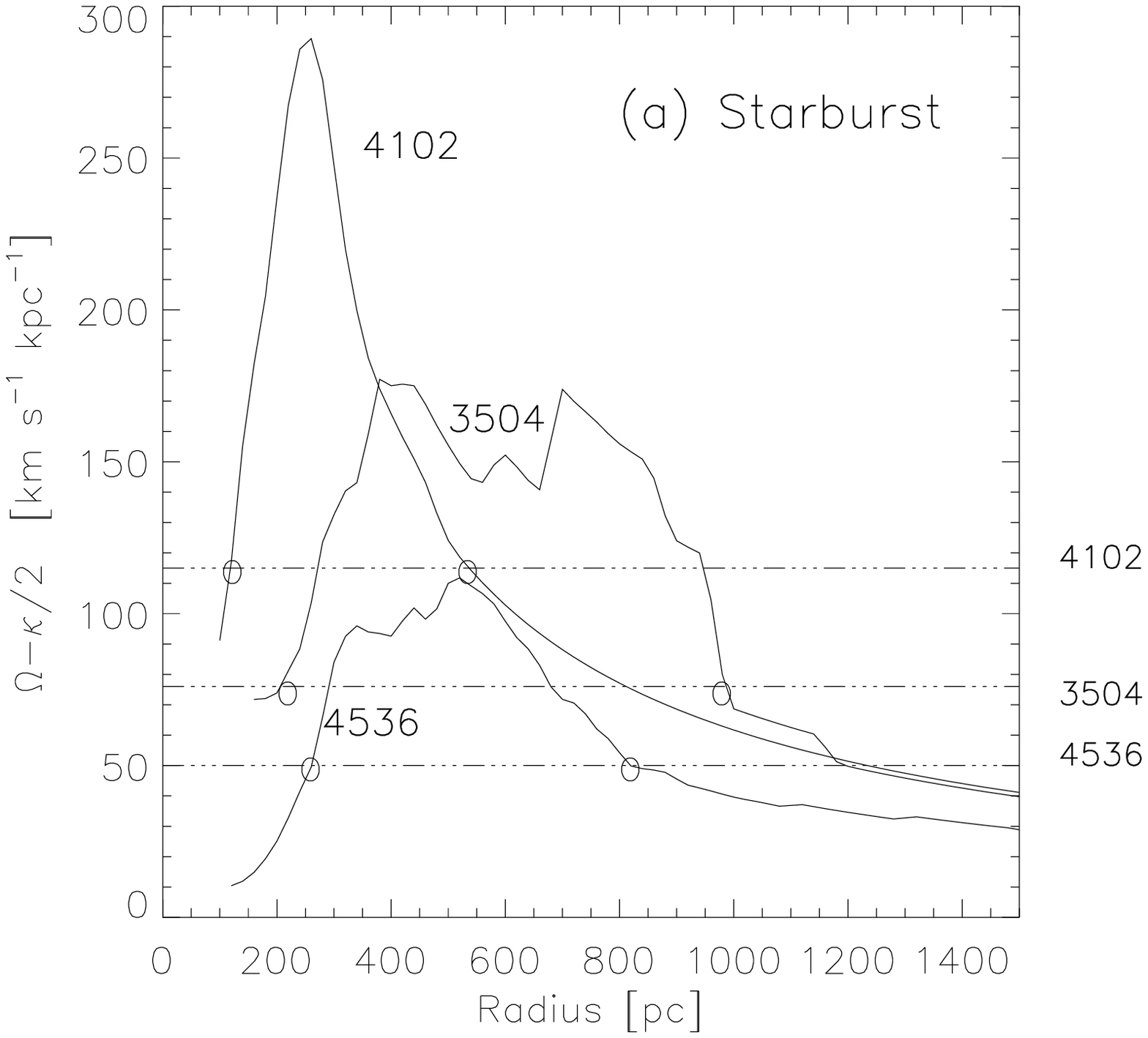}
\includegraphics[width=2.0in]{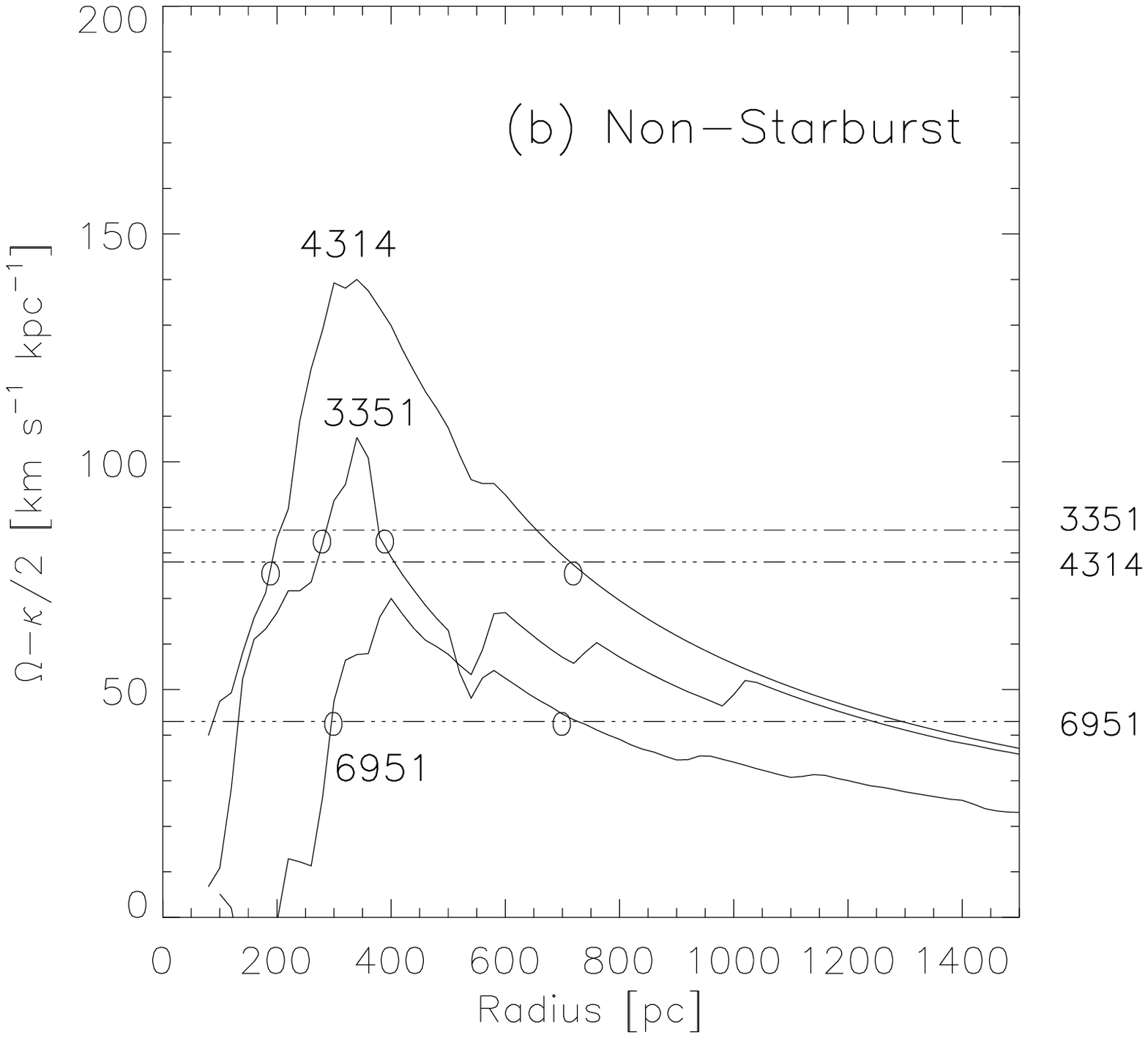}
\end{center}
\caption[]{
($\Omega - \kappa/2$) is plotted against radius.
The bar pattern speed $\Omega_{\mathrm{p}}$  is drawn as 
horizontal  lines and estimated  by assuming that the corotation 
resonance is  near the end of the bar.  
Under the epicycle theory for a weak bar, the intersection of
($\Omega - \kappa/2$) with $\Omega_{\mathrm{p}}$ 
defines the  locations of the ILRs. 
}
\end{figure}

\section{Circumnuclear Star Formation Morphology\break and Efficiency}

The starbursts and  non-starbursts have  circumnuclear SFR 
of  3--11   and  0.1--2\,M$_\odot$\,yr$^{-1}$, respectively,   
from RC and Br$\gamma$ data.
The SFE is therefore not a simple function of molecular gas content.
The CO and  star formation morphology are shown in Fig.~2.
%
For a given  CO-to-H$_{\rm 2}$ conversion factor,
the starbursts have a larger peak gas surface density $\Sigma_{\rm gas-m}$ 
in the inner 500\,pc radius than non-starbursts  
with a similar circumnuclear gas content (Fig.~4a).  
In the starbursts, both $\Sigma_{\rm gas-m}$  and $\Sigma_{\mathrm{SFR}}$  
increase towards the inner 500\,pc radius 
(Fig.~4a--b). Over the region of intense SF in several starbursts, 
$\Sigma_{\rm gas-m}$  remains close to the
critical density ($\Sigma_{\rm crit}$) for the onset of gravitational 
instabilities \cite{T64}, despite an order of magnitude variation in 
$\Sigma_{\rm crit}$  (Fig.~5b and e). 
%
In the non-starbursts, there are gas-rich regions with no 
appreciable  star formation, for instance, inside the 
ring of HII regions in NGC\,3351 and NGC\,4314, at the
CO peaks in NGC\,6951, and in the extended gas 
in NGC\,4569 (Fig.~2). 
The gas surface density, although high,  is still 
sub-critical in regions of inhibited star formation, 
as illustrated for NGC\,4314 in Fig.~5e--f.  
In NGC\,4569, the large local shear in the extended gas 
with large non-circular kinematics along 
the large-scale stellar bar may inhibit star formation.
I suggest  that circumnuclear starbursts 
produce a  high SFE by developing  supercritical  surface 
densities in a large fraction of the gas close to the center, 
while sub-critical densities and  large local shear 
may limit the SFE of non-starbursts. 
\begin{figure}[h]
\begin{center}
\includegraphics[width=3.5in]{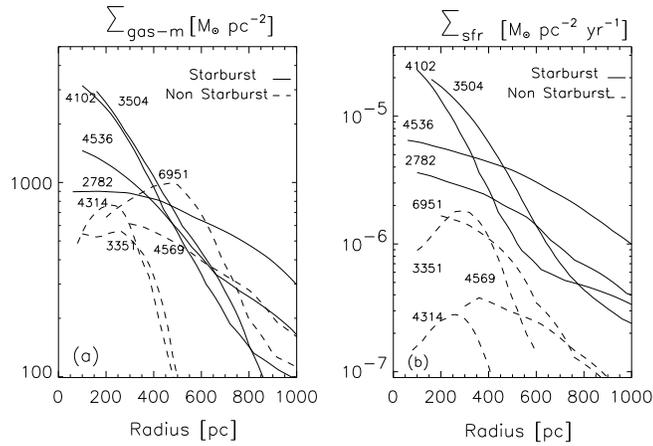}
\end{center}
\caption[]
{
\bf 
(a), (b) 
\rm
show the azimuthally averaged  molecular gas 
surface density ($\Sigma_{\mathrm{gas-m}}$) 
and SFR per unit area ($\Sigma_{\mathrm{SFR}}$). 
The  extinction-corrected $\Sigma_{\mathrm{SFR}}$ profiles are  
convolved to a similar resolution of 100--200\,pc  for all the galaxies
}
\end{figure}

\begin{figure}[h]
\begin{center}
\includegraphics[height=2.8in]{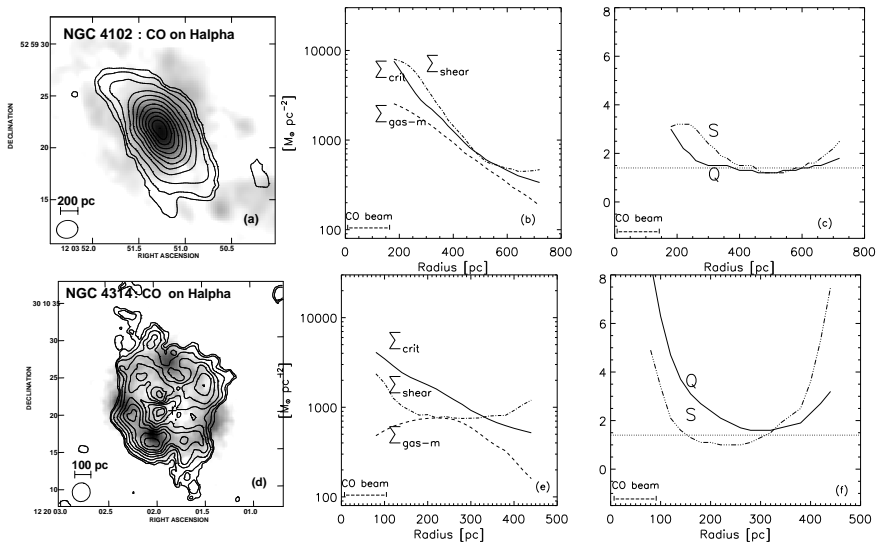}
\end{center}
\caption[]{
\small
\rm
{\bf (a, d)} the  CO  distribution  (contours) on 
the  H$\alpha$ (greyscale).  {\bf (b, e)} $\Sigma_{\rm gas-m}$, 
$\Sigma_{\mathrm{crit}}$, and  $\Sigma_{\mathrm{shear}}$. 
{\bf (c, f)} the Toomre Q 
and shear S parameters \cite{J99}.  
Quantities are plotted starting at a radius equal to the 
CO beam  size ($\sim$2$''$). 
In the non-starburst NGC\,4314, Q reaches its lowest value (1--2) in 
the ring  of HII regions  between r\,=\,250--400\,pc 
while at lower radii  where  there are no HII regions, Q increases to 6, 
indicating sub-critical gas densities.  
In the starburst NGC\,4102, 
Q remains  $\sim$1--2 between a radius of 250--700\,pc, 
over the region of intense star formation,  although  
$\Sigma_{\rm crit}$ varies by roughly an 
order of magnitude. 
\normalsize
}
\end{figure}

\section{The Extreme  Molecular Environment in the Inner Kpc}

\begin{table}
\caption{Molecular Gas Properties in the Circumnuclear Region} 
\begin{center}
\renewcommand{\arraystretch}{1.4}
\setlength\tabcolsep{5pt}
\begin{tabular}{lccc}
 {Quantities   }   &  {Outer Disk}      &
 {Inner r\,=\,500\,pc}      &  {Inner r\,=\,500\,pc} \\
 {}   &  { of Sa--Sc}      &
 {of sample galaxies }          &  {of Arp\,220}\\ 
\hline
(1) M$_{\rm gas,m}$ (M$_\odot$) & 
$\le$\,few\,$\times$\,$10^{9}$ & 
few\,$\times$\,(10$^{8}$--10$^{9}$)   & 
$ 3 \times 10^{9}$     \\
(2) M$_{\mathrm{gas}}$/M$_{\mathrm{dyn}}$  (\%) & 
$<$\,5    & 
10--30      &  
40--80      \\
(3) SFR (M$_\odot$~yr$^{-1}$) & ---  & 
0.1--11 
&  $>$\,100  \\
(4) $\Sigma_{\mathrm{gas-m}}$ (M$_\odot$~pc$^{-2}$) & 
1--100 & 
500--3500    &  
$4 \times 10^{4}$   \\
(5) $\sigma$ (km~s$^{-1}$) & 
6--10 & 
10--40
& 90  \\
(6) $\kappa$ (km~s$^{-1}$~kpc$^{-1}$) & 
$<$\,100 & 
800--3000   &  
$>$\,1000   \\
(7) $\Sigma_{\mathrm{crit}}$ (M$_\odot$~pc$^{-2}$)& 
$<$\,10 & 500--1500   & 2200  \\
(8) t$_{\mathrm{GI}}$ (Myr)& 
$>$\,10    & 0.5--1.5    & 0.5   \\
(9) $\lambda_{\mathrm{J}}$ (pc)&
few\,$\times$\,100--1000    & 100--300   & 90  \\
\hline
\end{tabular}
\end{center}
\label{apptab1b}
\small
The rows are : 
(1) molecular gas mass
(2) ratio of  molecular gas mass 
to dynamical mass;
(3) star formation rate; 
(4) molecular gas surface density;
(5) gas velocity dispersion; 
(6) critical Toomre 
density for the onset of gravitational instabilities
(7) epicyclic frequency; 
(8) growth  
timescale of the most unstable wavelength (Q/$\kappa$)
(9) Jeans length 
\normalsize
\end{table} 

\rm
Table 1 illustrates how molecular gas in the inner kpc and the 
outer disk differ markedly.  This has important implications 
for the circumnuclear  region. 
First, the high  molecular gas density
(several~100--1000\,M$_\odot$~pc$^{-2}$) 
and mass fraction (10--30\%) 
will lead to  enhanced self-gravity and clumpiness  of the gas 
 (e.g., \cite{SFB89}). 
The  two-fluid disk of gravitationally coupled gas and  stars
will be more unstable 
to gravitational instabilities than  a purely  stellar 
disk (e.g.,  \cite{J96}).
Second, in the presence of a large epicyclic frequency  
(several~100--1000\,km~s$^{-1}$~kpc$^{-1}$) and velocity dispersion 
(10--40\,km~s$^{-1}$),  gravitational instabilities can 
overcome Coriolis and pressure forces only at very high gas 
densities (few~100--1000\,M$_\odot$~pc$^{-2}$).
However, once triggered, they now grow on a timescale  
($t_{\rm GI}$)  as short as  a few Myrs, comparable to the 
lifetime of an OB star. 
These conditions  can enhance the fraction of gas converted into 
stars before a molecular cloud is disrupted by massive stars. 
Third, a high pressure, high turbulence ISM 
may  favor  more massive clusters (e.g., \cite{E93}) and 
it is relevant that 
many  sample galaxies show super star clusters in the inner 
kpc.  
Fourth, a comparison of Columns~3 and 4 in Table~1 suggests that  the 
prototypical ultra luminous infrared galaxy (ULIRG) Arp\,220  
may be  a scaled-up version of the starbursts in our sample. 
ULIRGs may be starbursts which have built an extreme 
molecular environment (density and linewidths) in the 
inner few 100 pc of a deep stellar potential well 
through major mergers or interactions.

\end{document}